\documentclass{nature}
\usepackage{bbm}
\usepackage{}
\usepackage{amssymb}
\usepackage{amsmath}
\usepackage{amsfonts}
\usepackage{graphicx}
\usepackage{caption}
\usepackage{subfig}
\usepackage{txfonts}
\usepackage{epsfig}
\begin{document}

\title{Controlling quantum spin Hall state via strain in various stacking bilayer phosphorene}
\author{Tian Zhang$^{1,2}$, Jia-He Lin$^{1,2}$, Yan-Mei Yu$^{1}$$^\star$ , Xiang-Rong Chen$^{2}$$^\star$ and Wu-Ming Liu$^{1}$$^\star$}

\maketitle

\begin{affiliations}
\item
Beijing National Laboratory for Condensed Matter Physics, Institute
of Physics, Chinese Academy of Sciences, Beijing 100190, China

$^{2}$Institute of Atomic and Molecular Physics, College of Physical
Science and Technology, Key Laboratory of High Energy Density
Physics and Technology of Ministry of Education, Sichuan University,
Chengdu 610065, China

$^\star$e-mail:ymyu@iphy.ac.cn;xrchen@scu.edu.cn;wliu@iphy.ac.cn

\end{affiliations}

\begin{abstract}
Quantum spin Hall (QSH) state of matter has a charge excitation bulk
bandgap and a pair of gapless spin-filtered edge-states, which can
support backscattering-free transport. Bilayer phosphorene possesses
a large tunable bandgap and high carrier mobilities, and therefore
has the widely potential applications in nanoelectronics and optics.
Here, we demonstrate an strain-induced electronic topological phase
transition from a normal to QSH state in bilayer phosphorene
accompanying by a band inversion that changes $\mathbbm{Z}_{2}$ from
0 to 1, which is highly dependent on the interlayer stacking. When
the bottom layer is shifted by 1/2 unit cell along axial direction
with respect to the top layer, the topological bandgap reaches up to
92.5 meV, which is sufficiently large to realize the QSH effect at
room temperature. Its optical absorption spectrum becomes broadened,
and even extends to the far-infra-red region leading to a wider
range of brightness, which is highly desirable in optic devices.

\end{abstract}

Two dimensional (2D) topological insulator (TI), namely, the quantum
spin Hall (QSH) insulator, has charge excitation bulk energy gap and
a pair of gapless spin-filtered edge states with a Dirac-cone-like
linear energy dispersion\cite{LED}. The special edge states are
topologically protected by the time reversal (TR) symmetry and can
immune to nonmagnetic scattering and geometry perturbations, thus
open new ways for backscattering-free transport. Such systems have
stimulated enormous research activity in condensed matter physics
due to their novel quantum spin Hall effect and hence the potential
application in quantum computation and spintronics\cite{APP1,APP2}.

Phosphorene, a new two-dimensional (2D) material, has been
successfully isolated by mechanical exfoliation \cite{BME} and
gained rapidly attention \cite{FNA,hcm1,RA1,RA2,RA3,RA4,RA5,RA6}.
Unlike graphene, phosphorene takes a puckered non-planar structure
although it is composed of the basic hexagon as shown in Fig. 1(a).
Phosphorene is chemically inert and has great transport properties
with high carrier mobilities (around $1000$ cm$^{2}$/V$\cdot$s) and
drain current modulation (four orders of magnitude larger than that
in graphene) \cite{FNA,hcm1,hcm2}, which makes the phosphorene a
potential candidate for future nanoelectronic applications
\cite{FNA}. Furthermore, phosphorene has been predicted as an ideal
direct band-gap material at the $\Gamma$ point of the first
Brillouin zone \cite{BZ1,RA2}, which is very important for
electronic and optical application.  A large number of 2D buckled
honeycomb structures have been predicted in nontrivial QSH states by
the first-principle calculations, including bilayers of group III
elements with Bi \cite{BL1}, 2D stable dumbbell stanene\cite{D1},
and so on \cite{BL1,D1,SO1,SO2}. It is reasonable to ask whether or
not a trival insulator phosphorene can be a topological nontrivial
insulator, which maybe largely improve its application on optical
and transport properties. Manifestation of the nontrivial topology
of occupied bands in a TI is attributed to band inversion (BI)
between occupied and unoccupied bands at the time reversal invariant
momenta (TRIM) in the bulk Brillouin zone by large enough spin-orbit
coupling \cite{SOC1,SOC2}, or external strain \cite{TW2,TW3}. There
are two common approaches to create or annihilate the TI states
\cite{TT}: (1) To undertake chemical tuning but often entails
uncontrolled effect of chemical disorder, (2) To change the band
topology of a physical system by adjusting lattice constants or
internal atomic positions in the unit cell \cite{TW1,TW2}, which is
an accepted way for phosphorene due to its highly sensitive of band
gap to external strain. Different from monolayer phosphorene,
few-layer phosphorene with weak interlayer van der Waals ($vdW$)
interaction has various interlayer stacking orders, which can
provide an exciting approach to tuning their optical and electronic
properties. Further, few-layer phosphorene has the smaller tunable
bandgap by external strain, and thus it was the more likely to
exhibit band inversion and the ensuing topological insulator
behavior. And bilayer phosphorene is the thinnest multilayer system,
which can provide fundamental information on the interlayer
interaction, and the stacking-dependent electronic and optical
properties, a feature akin to the bilayer graphene systems
\cite{bgs1,bgs2,bgs3}.

In this work, we explore the possibility of converting normal
insulating state into QSH one in bilayer phosphorene, through a
systematic investigation of electronic structure on interlayer
stacking and external strain. Our results reveal that the tuning of
topological behavior in bilayer phosphorene is highly dependent on
interlayer stacking order and the direction of the applied in-plane
strain, i.e., the direction with the maximum atomic wavefunction
overlap. Meanwhile, reversible direct to indirect and semiconductor
to metal transitions can be found by adjusting the weight of $vdW$.
The optical property of the QSH in bilayer phosphorene is examined
based on a real-space and real-time time-dependent density
functional theory, which proves that the absorption spectrum becomes
broadened, and even extends to the far-infra-red region and leads to
a wider range of brightness. Such improvement in optical responds
property is highly desirable in broadband photodetection and
infrared detector.

\section*{Results}
\subsection{Crystal structures and electronic structures of bilayer phosphorene for four different stacking orders.}
Bulk black phosphorus belongs to the space group $Cmca$, and
phosphorene can been viewed as cleaved from the (0001) surface of
black phosphorus. Bilayer phosphorene is held together by weak
interlayer forces with significant van der Waals character, which
allows formation of several polytypes due to the possibility of
different stacking. In this study we consider four possible
high-symmetry stacking orders: (1) the top layer is stacked
vertically on the bottom layer in the $Pmna$ (Ta), (2) the bottom
layer is shifted by half of one unit cell along $x$ or $y$ with
respect to the top layer in the $Pbcm$ (Tb), (3) the bottom layer is
shifted by one unit cell along $x$ or $y$ direction with respect to
the top layer, and thus the top and bottom layers are mirror images
of each other in the $Pmma$ (Tc), (4) the bottom layer is shifted by
one and a half of one unit cell along $x$ or $y$ direction with
respect to the top layer in the $Pccm$ (Td), as shown in Fig.
1(c)-1(f), respectively. Table 1 gives the optimized lattice
constants, bond length and other structural parameters of bilayer
phosphorene for the four stacking orders, which are well consistent
with the previously theoretical data\cite{JDXCZ} with errors lower
than $0.5\%$. For the different stacking order, the bond length
$R_{1}$ is almost same as the bond length $R^{'}_{1}$ but always
shorter than the bond length $R_{2}$. The bond angles $\alpha$ is
shorter than the $\beta$. The smallest layer interval $d_{int}$ in
the vertical direction is 3.503 {\AA} (Ta), 3.085 {\AA} (Tb), 3.739
{\AA} (Tc) and 3.291 {\AA} (Td). Our total energy calculations
indicate that the Ta-, Tb-, Tc- and Td-stacked orders have same
structural stability with the difference of the cohesive energy
below 10 meV. (The stability of Tb-stacked order is proved in the
supplementary information, which has the lowest cohesive energy.)

Fig. 2 gives the electronic structures for the four stacking orders.
The band gaps at $\Gamma$ are  0.434 eV (Ta), 0.442 eV (Tb), 0.264
eV (Tc) and 0.002 eV (Td). Since our HSE06 band structures have
indicated that these PBE band gaps are underestimated by $\sim$ 0.56
eV (see Supplementary Fig. S2), thus all four stacking orders should
have a band gap in excess of 0.56 eV, which agrees with the
previously theoretical data using the same method\cite{JDXCZ}. The
iso-surfaces of the band-decomposed charge density at $\Gamma$ point
of the valence band maximum (VBM) and conduction band minimum (CBM)
for the four stacking orders are shown in Fig. 2(e)-(h),
respectively, which show very similar charge density distribution,
and thus they have very similar bonding character. A prominent
in-plane anti-bonding-like feature of VBM (marked by blue dashed
box) and in-plane bonding-like features of CBM (marked by red dashed
box) are shown for each four stacking order. The similar out-plane
bonding and out-plane antibonding features are also found in the CBM
and VBM, respectively. We observe the overlap of layer-layer charge
density of CBM in Tc- and Td-stacked orders, showing the
bonding-like character, which is absent in Ta- and Tb-stacked
orders. We can expect that VBM and CBM are mainly originated from
the localized and delocalized states of P atoms, respectively, in
the interfacial area between the top and bottom layers. Hence, the
different interaction strength and band gaps are related to the
different $\pi$-$\pi$ interaction distance between the delocalized
states, which due to the different stacking order.

\subsection{x-y plane strain-induced quantum spin Hall state in bilayer phosphorene.}
The band gap $E_{g}$ changes with the in-plane strain $\sigma$ of
bilayer phosphorene for four stacking orders are shown in Fig. 3. We
consider three types of in-plane strains $\sigma_{x}$, $\sigma_{y}$
and $\sigma_{xy}$ along x direction, y direction, and both x and y
directions, respectively. When the strain is applied in given
direction, the lattice constants in the other directions are relaxed
fully through the technique of energy minimization in order to
ensure that the force on atoms is less than 0.01 eV/{\AA}. Our
calculation shows that all four orders are highly anisotropic. Band
gaps of four stacking orders are highly sensitive to in-plane
strain, where the band gap is more sensitive to $\sigma_{xy}$ as
compared with those of strains $\sigma_{x}$, $\sigma_{y}$. Hence,
for the remainder of this study, we mainly focus on the effect of
$\sigma_{xy}$ on band gap in Fig. 3. When $\sigma_{xy}$$>$0
(tensile), we can find the band gaps increase with $\sigma_{xy}$
increasing, then turn to decrease at $\sigma_{xy}$ = $4.0 \%$ (Ta)
and $\sigma_{xy}$ = $3.0 \%$ (Tb). When $\sigma_{xy}$$<$0
(compression), or near zero strain for Td-stacked order, the band
gap decrease with $\sigma_{xy}$ increasing, then turn to increase at
$\sigma_{xy}$ = -3.0$\%$ (Ta and Tb), and $\sigma_{xy}$ = $0.02\%$
(Tc). This corresponding BI process, as shown in inset in Fig. 3(d),
indicates a possible topological phase transition. The
direct-indirect band gap transition can been found when
$\sigma_{xy}$=$-2.0 \%$ in Tc- and Td-stacked orders. With the
further compression strain increasing, The band gaps are closed and
turn into metal states when $\sigma_{xy}$=$-7.0\%$ (Tb),
$\sigma_{xy}$=$-3.0\%$ (Tc), and $\sigma_{xy}$=$-5.0\%$ (Td).

In Fig. 4(a)-(c), we further analyze the electronic band structures
for Tb-stacked order at $\sigma_{xy}$ values. We note that, as
$\sigma_{xy}$ increases, its VBM and CBM start approaching each
other and then overlapping when $\sigma_{xy}$=$\sim -2.77\%$ . For
stronger compression strain ($\sigma_{xy}$=$-3.0\%$), repulsion
between the electronic bands leads to shift VBM at $\Gamma$ and CBM
at $\Gamma$ away from each and enlarges the band gap, accompanied by
the inversion of the top of valence and bottom of the conduction
band at the $\Gamma$ point. The band parity also change its sign due
to BI. Such band-inversion character is also observed in the density
of states (DOS) and the orbital-projected band structures when
$\sigma_{xy}$=$-3.0\%$ [see Fig. 4(e)-(g)]. The $P-p$ makes a
significant contribution to the total DOS, and the conduction band
near $E_{F}$ is from the $p_{z}$ and $p_{y}$ obitals, while the
valence band around $E_{F}$ is mainly contributed by the $p_{z}$
orbital. By observing the orbital-projected band structures under
$\sigma_{xy}$=$-3.0\%$, we find that the weight of $p_{z}$ orbital
is very large at the local region of CBM around $\Gamma$ point. The
weight of $p_{y}$ orbital is significantly large at the local region
of VBM around $\Gamma$ point . This phenomenon shows a obvious BI
process when the compression strain $\sigma_{xy}$ is increased.

Here, we use a rigorous method of Fu and Kane \cite{FK} to prove
that Tb-stacked order when $\sigma_{xy}$=$0$ is a topologically
trivial band insulator with $\mathbb{Z}_{2}$=$0$, while for larger
compression strain $\varepsilon_{xy}$=-3.0$\%$, it is in a
nontrivial QSH state with $\mathbbm{Z}_{2}$=$1$. The method is valid
since the Tb-stacked bilayer has both spatial invention and time
reversal symmetries (four time reversal invariant points in the 2D
Brillouin zone). Inversion center in the crystal ensures
$\varepsilon_{n\alpha}(k)=\varepsilon_{n\alpha}(-k)$, where
$\varepsilon_{n\alpha}(k)$ is the electron energy for the $n$-th
band with spin index $\alpha$ at k wave vector in the Brillouin
zone. The time reversal symmetry makes
$\varepsilon_{n\alpha}(k)=\varepsilon_{n\bar{\alpha}}(-k)$, where
$\bar{\alpha}$ is the spin opposite to $\alpha$. The calculated
parities of all occupied bands at four time-reversal invariant
momenta are listed in Table 2. We can find the product of parities
of occupied bands contributes a +1 at the four time-reversal
invariant momenta when $\sigma_{xy}$=$0$, yielding a trivial
topological invariant $\mathbbm{Z}_{2}$=$0$. As the strain is
increased up to $\sigma_{xy}$=$-3.0\%$, band inversion at the
$\Gamma$ point takes place. The product of parities of occupied
bands is -1 at $\Gamma$ while +1 at the three other time-reversal
invariant momenta. Thus the Tb-stacked order under the compression
strain $\sigma_{xy}$=$-3.0\%$ are identified as topological
insulators with $\mathbbm{Z}_{2}$=1. The results shown in Fig. 4(d)
suggest that the QSH state in Tb-stacked order survives under the
compression strain $\sigma_{xy}$ from $-2.77 \%$ to $-7.0 \%$, where
the maximum band gap $E_{g}$$=$92.5 meV can be found when
$\sigma_{xy}$=$-5.0 \%$. (The calculated critical strain
$\sigma_{xy}$=$-5\%$ where Tb-stacked order is in quantum spin Hall
state by HSE06 method, which is presented in Supplementary Fig.S3).

We do not observe the topological phase transition in Tc-stacked
order because BI do not appears by adjusting in-plane strain. We
also do not find topological phase transition in Td-stacked order
because the VBM and the CBM have the same parity, which is not
closely relate to the details of the atoms orbitals or bond types
but implies different signs of overlap integral of the atomic
orbitals. The detailed topological phase transition process for
Ta-stacked order is given in supplementary information.

\subsection{Universal reversible semiconductor-metal transition by interlayer interaction.}
We have employed a semi-empirical van der Waals ($vdW$) approach, as
proposed by Grimme known as the DFT-D2 method to correctly describe
the interlayer interaction between the top and bottom layers of
bilayer phosphorene. The total energy of system is defined as
$E_{DFT-D}$=$E_{KS-DFT}$+$E_{vdW}$, where $E_{KS-DFT}$ is the
conventional Kohn-Sham DFT total energy and $E_{vdW}$ the total
energy can be described via a simple pair-wise force field which is
optimized for several popular DFT functionals. To focus on the
effect of interlayer interaction in electronic structure, a scaling
factor is added in the front of the $vdW$ term in the computations,
referring as the weight of $vdW$ ($WvdW$).

The dependence of the band gaps on $WvdW$ for four different
stacking orders are presented in the Fig. 5(a). As compared with the
monolayer phosphorene, the band gap of the bilayer phosphorene is
smaller and each band becomes doubly degenerate at $WvdW$=$0t$
($t$=1 is the energy unit) as shown in Fig. 5(b) and (c), indicating
the absence of chemical interaction between the top and bottom
layers in Td-stacked order. As $WvdW$ increases, the double
degeneracy of the band is broken because the top and bottom layers
start to bond together. The change of the band gap on $WvdW$ is more
strongly in Td-stacked order, compared with that of Ta-, Tb- and
Tc-stacked orders. As shown the inset in Fig. 5(a), the band gap is
closed when $WvdW$=$\sim$ 1.05 $t$, then opens up again for the
larger $WvdW$, which indicates a BI process. The weight of $p_{z}$
in CBM has a clear saltation when $WvdW$=$1.14 t$, compared with
that when $WvdW$=$0 t$ as shown in Fig. 5(b) and (c), proving BI of
VBM and CBM. However, Td-stacked bilayer phosphorene can not be the
topological insulator by adjusting $WvdW$ because the parities of
its VBM and CBM are the same. With the $wvdW$ further increasing
beyond 2 $t$, the band gaps of all four structures are closed due to
overlap of VBM and CBM, indicating that the four structures undergo
semiconductor to metal transition. In the whole process, their band
gaps have the large variation, but we do not observe any sign of
topological phase transition, indicating topological phase
transition can not be induced by $WvdW$. Hence, we can expect that
the topological phase transition is closely relationship with the
direction of the applied strain, which would be the one along
bonding direction, i.e., the direction with maximum atomic wave
function overlap.

\subsection{Optical responds of Tb-stacked bilayer phosphorene under the $x-y$ plane train $\sigma_{xy}$.}
According to the above discussion, we know the Tb-stacked bilayer
phosphorene is in a nontrivial state when the compression strain
$\sigma_{xy}=3.0 \%$, and have verified its stability by calculating
the vibration spectra. The detailed analysis of vibration spectra is
shown in the Supplemental information. As we know, phosphorene has
promising optical properties. Here, we mainly study the optical
responds of Tb-stacked bilayer phosphorene when $\sigma_{xy}=-3.0
\%$ in order to make a compare.

As shown in Fig. 6 (b), the obtained photonic band gap (PBG) are 0.5
eV and 0.01 ev, respectively, when $\sigma_{xy}=-3.0\%$ and
$\sigma_{xy}=0$. The PBG is close to the band gap calculated by
VASP, which means that OCTOPUS is reliable on calculation for the
optical property. When the impulse excitation polarizes in the
armchair-edge direction ($y$-direction) or the zigzag-edge direction
($x$-direction), there is always a main absorption peak around 10 eV
, because there is only a $\sigma$ plasmon resonance mode in bilayer
phosphorene. When the impulse excitation polarizes in the
zigzag-edge direction, the main absorption peak when
$\sigma_{xy}=-3.0\%$ becomes smaller and shows blue shift comparing
to that when $\sigma_{xy}=0$. This is because the resonance level
spacing increases with the interatomic spacing increasing in the
zigzag-edge direction. When the impulse excitation polarizes in the
armchair-edge direction, the main absorption peak around 10 eV when
$\sigma_{xy}=-3.0\%$ is almost unchange compared to that when
$\sigma_{xy}=0$. In the low-energy resonance zone from 0 eV to 2 eV,
their strength of optical absorption for both $\sigma_{xy}=-3.0\%$
and $\sigma_{xy}=0$ are nearly zero, when impulae excitation
polarizes in the zigzag-edge direction. Meanwhile, the optical
absorption are highly active in the energy band gap of bilayer
phosphorene, when the impulse excitation polarizes in the
armchair-edge direction. This phenomenon is the result of selection
rules associated with the anisotropic symmetries of bilayer
phosphorene. Interestingly, when the impulse excitation polarizes in
the armchair-edge direction, compared with that when
$\sigma_{xy}=0$, the absorption spectrum is red-shift. Furthermore,
we find a weak absorption closed to zero (0.01 eV) owing to the
decrease in energy gap when $\sigma_{xy}=-3.0\%$. To elucidate the
mechanism of the optical absorption in low-energy resonance behind
the resonance phenomena, the induced charge response has been
analyzed in real-time propagation, as shown in Fig. 6(c)-(f).

We analyze the induced charge density at the low-energy resonance
zone. Fig. 6(c)-(f) presents the Fourier transform of the induced
charge density when $\sigma_{xy}$=0 and $-3.0\%$ along the
armchair-edge direction and the zigzag-edge direction. We set the
induced density plane to parallel the bilayer phosphorene plane, and
to locate the middle of the top pucker-layer in the vertical
direction. When the impulse excitation polarizes in the zigzag-edge
direction, the induced electron density is separate from the induced
hole density roughly, and they locate at the left side and the right
side (seen in the black box and the white box), respectively, as
shown in the Fig. 6(c) at $\sigma_{xy}=0$ and 6(e) at
$\sigma_{xy}=-3.0\%$. Hence, we can find a common characteristic
that the induced charge at the most resonance points is distributed
at the boundary region in Fig. 6(c) and 6(e). The induced charge
density of $\sigma_{xy}=0$ is more plentiful than at
$\sigma_{xy}=-3.0\%$ by comparing Fig. 6(c) and 6(e). This
phenomenon supports that absorption spectrum of $\sigma_{xy}=-3.0\%$
becomes weaker compared to $\sigma_{xy}=0$. When the impulse
excitation polarizes in the armchair-edge direction as shown in Fig.
6(d) at $\sigma_{xy}=0$ and 6(f) at $\sigma_{xy}=-3.0\%$, the
induced electron density is also separate from the induced hole
density, but they locate at the up and down (seen the black box and
the white box), respectively, as shown in the Fig. 6(d) at
$\sigma_{xy}=0$ and 6(f) at $\sigma_{xy}=-3.0\%$. Furthermore, the
induced charge density of the $\sigma_{xy}=-3.0\%$ is richer and
concentrated in the center of the bilayer phosphorene when the
impulse excitation polarizes in the armchair-edge direction. What is
because that the shielding effect becomes stronger with the
interatomic spacing becoming smaller.

\section*{Discussion}
we demonstrate an in-plane strain-induced electronic topological
transition from a normal to QSH state in bilayer phosphorene,
accompanying by a band inversion that causes the change in the
$\mathbbm{Z}_{2}$ topological invariant from 0 to 1. Our investigate
shows that the topological phase transition in bilayer phosphorene
is closely related to interlayer stacking and the direction of
applying strain, which would be the one along bonding direction,
i.e., the direction with maximum atomic wave function overlap. The
topologically non-trivial bandgap in Tb-stacked bilayer phosphorene
can reach up to 92.5 meV when $\sigma_{xy}$=$-5.0 \%$, which is
sufficiently large to realize the QSH effect at room temperature.
Meanwhile, reversible direct to indirect and semiconductor to metal
transitions can be found by adjusting the weight of $vdW$. The
optical absorption spectrum of the QSH state in bilayer phosphorene
becomes broadened, and even extends to the far-infra-red region and
leads to a wider range of brightness, which is highly desirable in
broadband photodetection and infrared detector.

Note added in proof: during the review process for this manuscript,
we noted an theoretical work that in few-layer ($>$ 2) phosphorene
there will be a normal-to-topological phase transition induced
purely by applying an electric field\cite{ps}. However, the
normal-to-topological phase transition in various stacking bilayer
phosphorene has not yet been explored.

\section*{Methods}
\subsection{$\mathbbm{Z}_{2}$ calculation technique.}
We use the method of Fu and Kane\cite{FK} to calculate topological
invariant $\mathbbm{Z}_{2}$. $\mathcal {H}$ is a time-reversal
invariant periodic Hamiltonian with 2$N$ occupied bands
characterized by Bloch wave functions. A time-reversal operator
matrix relates time-reversed wave functions is defined by
\begin{equation}
    A_{\alpha\beta}(\Gamma_{i})=<\mu_{\alpha}(\Gamma_{i})|\Theta|\mu_{\beta}(\Gamma_{-i})>
\end{equation}
where $\alpha$, $\beta$ =1, 2, ..., $N$,
$|\mu_{\alpha}(\Gamma_{i})>$ are cell periodic eigenstates of the
Bloch Hamiltonian, $\Theta$=exp($i\pi$$S_{y})K$ is the time-reversal
operator ($S_{y}$ is spin and $K$ complex conjugation ), which
$\Theta^{2}$=-1 for spin 1/2 particles. Since
$<$$\Theta\mu_{\alpha}(\Gamma_{i})|\Theta\mu_{\beta}(\Gamma_{i})$$>$=$<$$\mu_{\beta}(\Gamma_{i})|\mu_{\alpha}(\Gamma_{i})$$>$,
$A(\Gamma_{i})$ is antisymmetric at TRIM $\Gamma_{i}$. The square of
its Pfaffian is equal to its determinant, i.e.,
det[$A$]=Pf$[A]^{2}$. Then
$\delta_{i}$=(det[$A$($\Gamma_{i}$)])$^{1/2}$/Pf$[A(\Gamma_{i})]$=$\pm
$1. Hence, the topological invariant $\mathbbm{Z}_{2}$ can be
defined as
\begin{equation}
    (-1)^{\mathbbm{Z}_{2}}=\prod^{4}_{i=1}\delta_{i}
\end{equation}
When solids have space-reversal symmetry, $\mathbbm{Z}_{2}$ can be
simplified as
\begin{equation}\label{22}
    (-1)^{\mathbbm{Z}_{2}}=\prod^{M}_{i=1}\xi_{2m}(\Gamma_{i})
\end{equation}
where $\xi$ is the parities of all occupied bands at $\Gamma_{i}$,
and $M$ is the number of the time-reversal invariant points.

\subsection{Electronic structure calculation technique.}
We calculate the lattice configurations as well as electronic band
structures of bilayer phosphorene with four different stacking
orders based on the density functional theory (DFT) implemented in
the Vienna Ab-initio Simulation Package (VASP)\cite{vasp1,vasp2}.
The projector augmented wave (PAW)\cite{PAW} method and the
Perdew-Eurk-Ernzerhof (PBE)\cite{PBE} exchange-correlation
functional are adopted. Long-range dispersion corrections have been
taken into account within a semi-empirical van der Waals approach
proposed by Grimme known as the DFT-D2 method \cite{DFTD2} (where D2
stands for the second generation of this method). The kinetic energy
cutoff for the plane wave basis set is chosen to be $600$ eV, and
the reciprocal space is meshed at $14\times 10\times 1$ using
Monkhorst-Pack method\cite{mpm}. A vacuum space of at least 25 {\AA}
along the $z$ direction is used to separate the bilayer systems in
order to avoid spurious interactions due to the nonlocal nature of
the correlation energy\cite{ce}. In order to correct the PBE band
gaps, we apply a hybrid Heyd-Scuseria-Emzerhof (HSE)\cite{HSE}
functional in which the exchange potential is separated into a
long-range and a short-range part, where $1/4$ of the PBE exchange
is replaced by the Hartree-Fock exact exchange and the full PBE
correlation energy is added. Hence the HSE functional is thought to
correct the GGA band gaps \cite{HSEB} significant by partially
correcting the self-interaction.

\subsection{Optical property calculation technique.}
We calculate the optical response of the Tb-stacked bilayer
phosphorene under the strains $\sigma_{xy}=0.0\%$ and
$\sigma_{xy}=-3.0\%$, based on a real-space and real-time
time-dependent density functional theory (TDDFT) as implemented in
the OCTOPUS code\cite{OCTOPUS}. The Hartwigsen-Goedecker-Hutter
pseudopotentials\cite{HGHP} and Generalized Gradient Approximation
(GGA) with PBE functional for the exchange-correlation are used to
calculate both the ground state and excited state. we mainly
investigated the plasmon excitation in the direction that is
parallel to the plane of the phosphorene. Geometries of phosphorene
which we mainly discussed is rectangular. The simulation zone was
defined by assigning a sphere around each atom with a radius of 6
{\AA} and a uniform mesh grid of 0.3 {\AA}. In the real time
propagation, excitation spectrum was extracted by Fourier transform
of the dipole strength induced by an impulse excitation\cite{iex}.
In the real-time propagation,the electronic wave packets were
evolved for typically 6000 steps with a time step of
$\Delta$$t$=0.005 {\AA}/eV.



\begin{addendum}
\item [Acknowledgement]
T. Zhang acknowledges very helpful discussions with M. Yang, H. J.
Lin and X. L. Zhang. This work was supported by the NKBRSFC under
grants Nos. 2011CB921502, 2012CB821305, 2010CB922904, NSFC under
grants Nos. 11228409, 61227902, 11174214, NSAF under grant No.
U1430117.

\item [Author Contributions]
T. Z. performed calculations. T. Z., J. H. L., Y. M. Y., X. R. C.,
W. M. L. analyzed numerical results. T. Z., J. H. L., Y. M. Y., X.
R. C., W. M. L. contributed in completing the paper.

\item [Competing Interests]
The authors declare that they have no competing financial interests.

\item [Correspondence]
Correspondence and requests for materials should be addressed to
Tian Zhang and Wu-Ming Liu.

\end{addendum}

\clearpage

\newpage
\bigskip
\textbf{Figure 1 Four stacking orders of bilayer phosphorene.} (a)
The monolayer black phosphorus (phosphorene), where the top and
bottom P atoms of the nonplanar sublayers are represented by green
and purple atoms. (b) The projection of monolayer black phosphorus
on $x$-$y$ plane, where $R_{1}$ and $R^{'}_{1}$ are two types of P-P
in-plane bond lengths, and $\alpha$ is the angle between two $R_{1}$
bonds. One unit cell has four atoms as included by the blue shadowed
region. Note that filled and opened dots with the same color are two
different sublattice. (c)-(f) The four different stacking orders:
(Ta) the top layer is stacked vertically on the bottom layer, (Tb)
the bottom layer is shifted by half of one unit cell along x or y
with respect to the top layer, (Tc) the bottom layer is shifted by
one unit cell along x or y direction with respect to the top layer,
and thus the top and bottom layers are mirror images of each other,
(Td) the bottom layer is shifted by one and a half of one unit cell
along x or y, where $R_{2}$ is the length of the out-of-plane bond,
$\beta$ is the angle between the in-plane and out-plane bonds and
$d_{int}$ is the smallest layer interval in the vertical direction.

\bigskip
\textbf{Figure 2 The electronic structures of bilayer phosphorene
for four different stacking orders.} (a)-(d) The band structures.
The maximum valence band and the minimum conduction band are
represented by purple and red lines, and valence band maximum (VBM)
and conduction band minimum (CBM) are denoted by purple and red real
boxes. The band gaps $E_{g}$ are indicated by blue shadowed regions.
The inset in (d) is the result of enlarging the region close to the
Fermi level. The Fermi level is set to zero. $\Gamma$ (0.0, 0.0,
0.0), X (0.0, 0.5, 0.0) and Y (0.5, 0.0, 0.0) refer to special
points in the first Brillouin zone. (e)-(h) The isosurface of
band-decomposed charge densities on y-z plane corresponding to VBM
and CBM at $\Gamma$ point, where the top and bottom P atoms of the
nonplanar sublayers are represented by green and purple atoms. The
bonding and antibonding features are highlighted by red and blue
dashed boxes. The isosurface in subfigures (e)-(h) is set to be
0.0037 e/{\AA}$^{3}$.

\bigskip
\textbf{Figure 3 The band gap $E_{g}$ changes with the in-plane
strain $\sigma$ of bilayer phosphorene for four stacking orders.}
(a)-(d) The band gaps for four different stacking orders on the
in-plane strain, where $\sigma_{x}$, $\sigma_{y}$ and $\sigma_{xy}$
mean the strain along x direction (zigzag direction), y direction
(armchair direction) and both x and y directions, respectively. We
consider tension strain ($\sigma$ $>$ 0), and compression strain
($\sigma$ $<$ 0). The direct-indirect band-gap transition can be
delineated by vertical blue dashed line, and the condition under
which bilayer phosphorene becomes metallic is highlighted by blue
shaded region. We find the band gap is reopened when in-plane strain
up to the critical value, as denoted by green dashed box. The inset
in (d) is the enlarged result of green dashed box.
\bigskip

\textbf{Figure 4 Electronic structures of bilayer phosphorene for
Tb-staked order.} (a)-(c) The band structures of bilayer phosphorene
for Tb-stacked order when the in-plane strain $\sigma_{xy}$ is
2.0$\%$, 2.77$\%$ and 3.0$\%$, where the maximum valence band and
minimum conduction band are represented by the purple and red lines,
and (even, odd) parity is denoted by (+, -). The inset in (c) is the
result of enlarging the region close to the Fermi level. The Fermi
level is set to zero. $\Gamma$ (0.0, 0.0, 0.0), X (0.0, 0.5, 0.0)
and Y (0.5, 0.0, 0.0) refer to special points in the first Brillouin
zone. (d) The valence band maximum (VBM) and conduction band minimum
(CBM) of bilayer phosphorene for Tb-stacked order changes with the
compression strain $\sigma_{xy}$, where the CBM and VBM are
represented by red and purple lines, the CBM and VBM at $\Gamma$ are
represented by red and purple dashed lines, and the VBM, CBM and CBM
along $\Gamma$ -X, $\Gamma$ -X and $\Gamma$ -Y are represented by
orange, green and blue dashed lines. The condition under which
Tb-stacked bilayer phosphorene becomes metallic is highlighted by
red shadowed region. (e) The density of states of Tb-stacked bilayer
phosphorene when $\sigma_{xy}=-3.0 \%$, where the total density of
states (DOS)is represented by the gray dotted lines, the $s$- and
$p$-orbitals of P atom are represented by the purple and orange
dotted lines, and the $p_{x}$, $p_{y}$ and $p_{z}$ orbitals are
represented by red and blue dotted lines. (f) and (g) The $p_{y}$
and $p_{z}$ orbital-projected band structures of Tb-stacked bilayer
phosphorene, where the radii of circles are proportional to the
weight of corresponding orbital.
\bigskip

\textbf{Figure 5 The electronic structures changes with $WvdW$ of
bilayer phosphorene for four stacking orders.} (a) The band gap
$E_{g}$ changes with $WvdW$, where the band gaps $E_{g}$ of Ta-,
Tb-, Tc- and Td-stacked bilayer phosphorene are represented by the
red, green, orange and blue points, and their fitted lines can be
denoted by the corresponding color-lines. We find the band gap of
Td-stacked bilayer phosphorene is reopened when $WvdW$ up to
critical value, as shown the inset in (a). (b) and (c) The $p_{z}$
orbital-projected band structures of Td-stacked bilayer phosphorene
when $WvdW=0$ and $WvdW=1.14t$, where the weights of $p_{z}$ orbital
are proportional to the radii of circles, and (even, odd) parity is
denoted by (+, -). The inset in (c) is the enlarged result of green
dashed box. The Fermi level is set to zero. $\Gamma$(0.0, 0.0, 0.0),
X(0.0, 0.5, 0.0) and Y(0.5, 0.0, 0.0) refer to special points in the
first Brillouin zone.
\bigskip

\textbf{Figure 6 The optical property of bilayer phosphorene for
Tb-stacked order.} (a) The top and side views of Tb-stacked bilayer
phosphorene, where up and bottom P atoms of the nonplanar sublayers
are represented by the purple and green atoms, and hydrogen atom is
represented by the red atom to passivate the dangling $\sigma$ bonds
at the edges. (b) The optical absorption of Tb-stacked bilayer
phosphorene when the compression strain $\sigma_{xy}=0.0\%$ and
$\sigma_{xy}=-3.0\%$. We consider two polarized directions, x
(zigzag) and y (armchair) directions. Red arrows point out the
moving direction of the absorption peaks after applying in-plane
strain $\sigma_{xy}$, as shown the inset in (b). (c) and (d) The
induced charge density of Tb-stacked bilayer phosphorene without
strain to an impulse excitation polarized in the $x$- and
$y$-directions. (e) and (f) The induced density of Tb-stacked
bilayer phosphorene when the compression strain $\sigma_{xy}=-3.0\%$
to an impulse excitation polarized in the $x$- and $y$-directions.
The regions where electron and hole density mainly locates, are
highlighted by black and white dashed boxes. The selected energy
resonance points are 4.96 eV (c), 1.11 eV (d), 4.96 eV (e) and 1.22
eV (f).

\newpage
\begin{figure}
    \begin{center}
        \epsfig{file=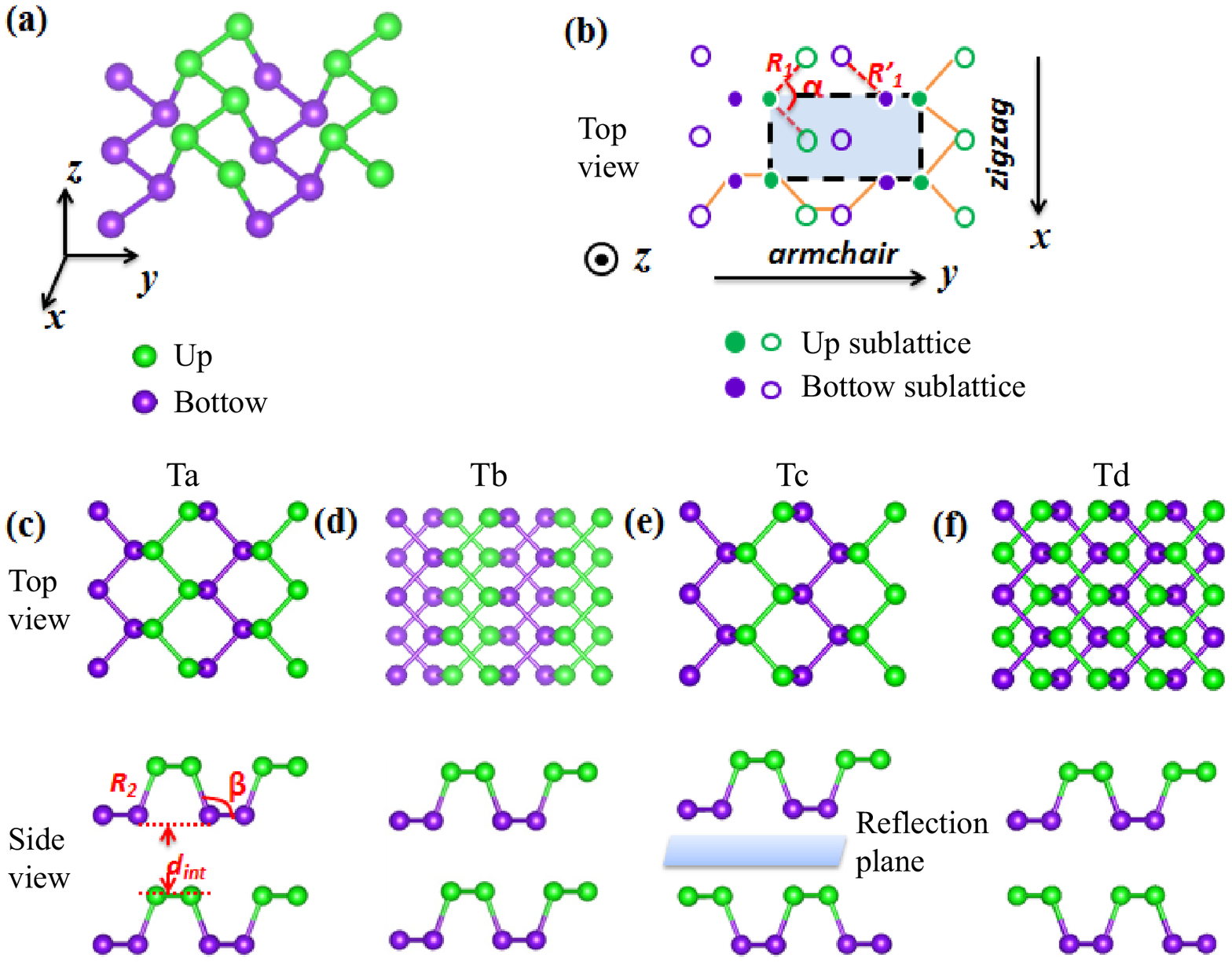,width=15cm}
    \end{center}
    \label{fig:Geo}
\end{figure}

\newpage
\begin{figure}
    \begin{center}
        \epsfig{file=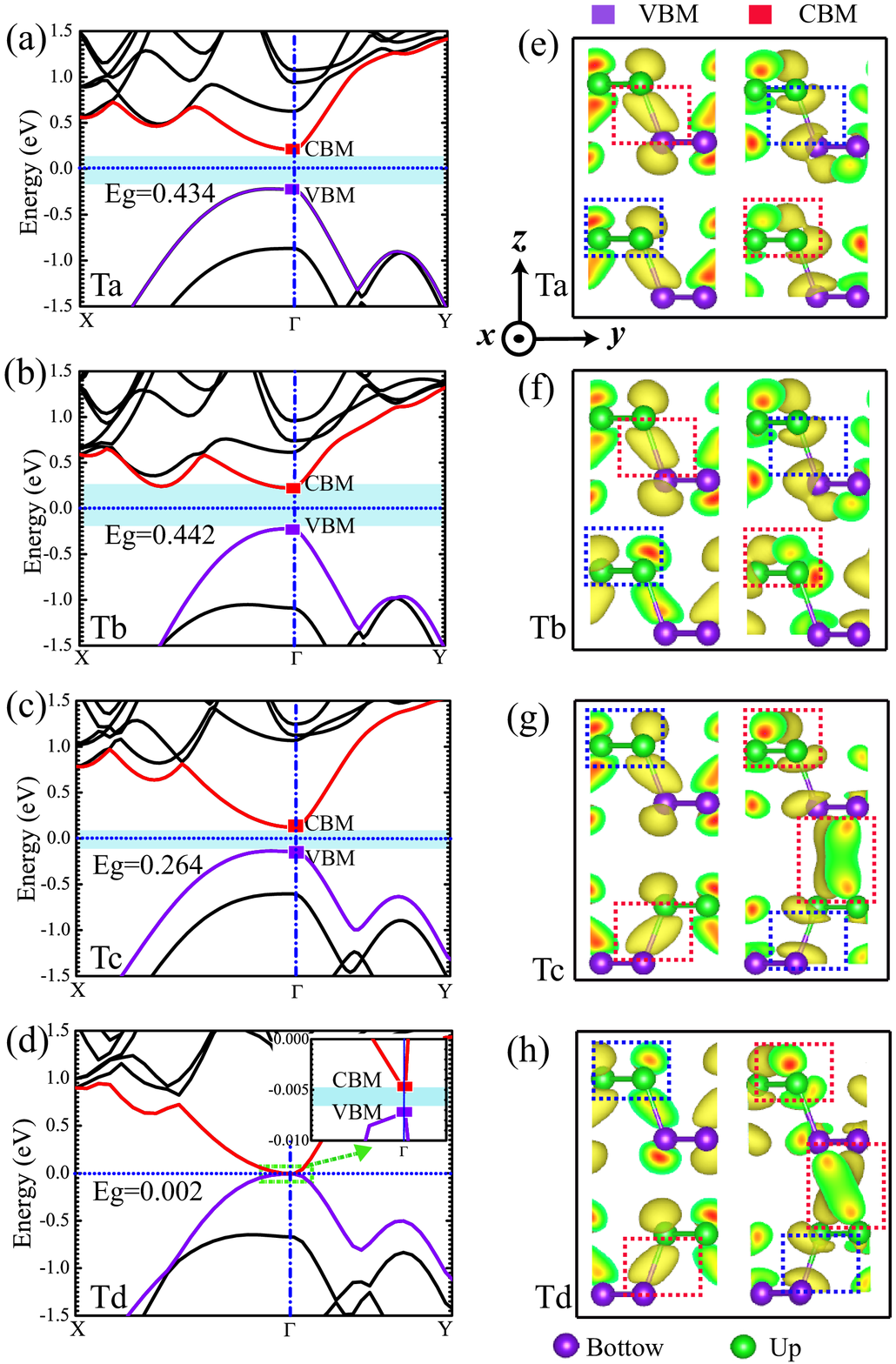,width=12cm}
    \end{center}
    \label{fig:DOS}
\end{figure}

\newpage
\begin{figure}
    \begin{center}
        \epsfig{file=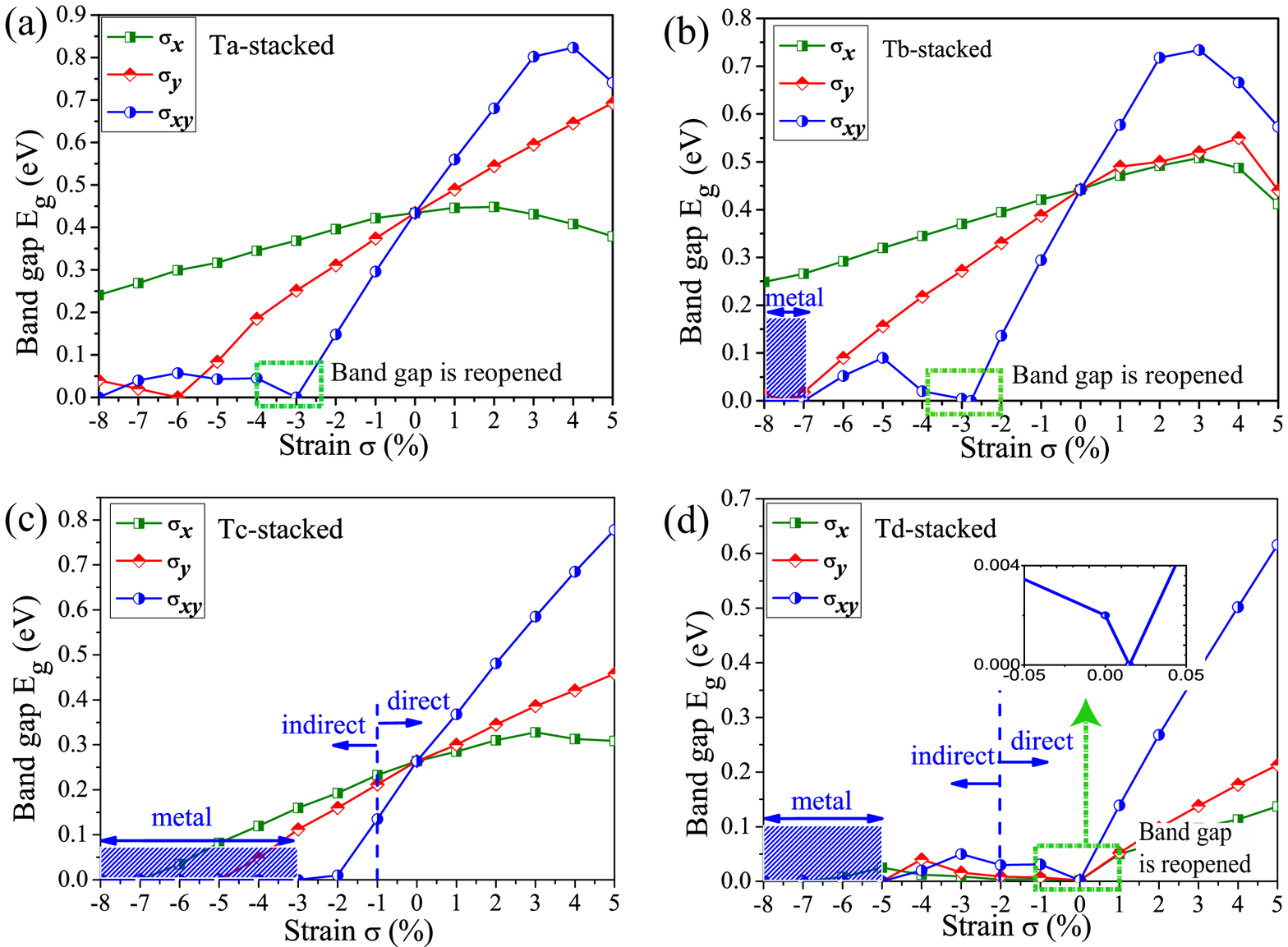,width=16cm}
    \end{center}
    \label{fig:TB}
\end{figure}

\newpage
\begin{figure}
    \begin{center}
        \epsfig{file=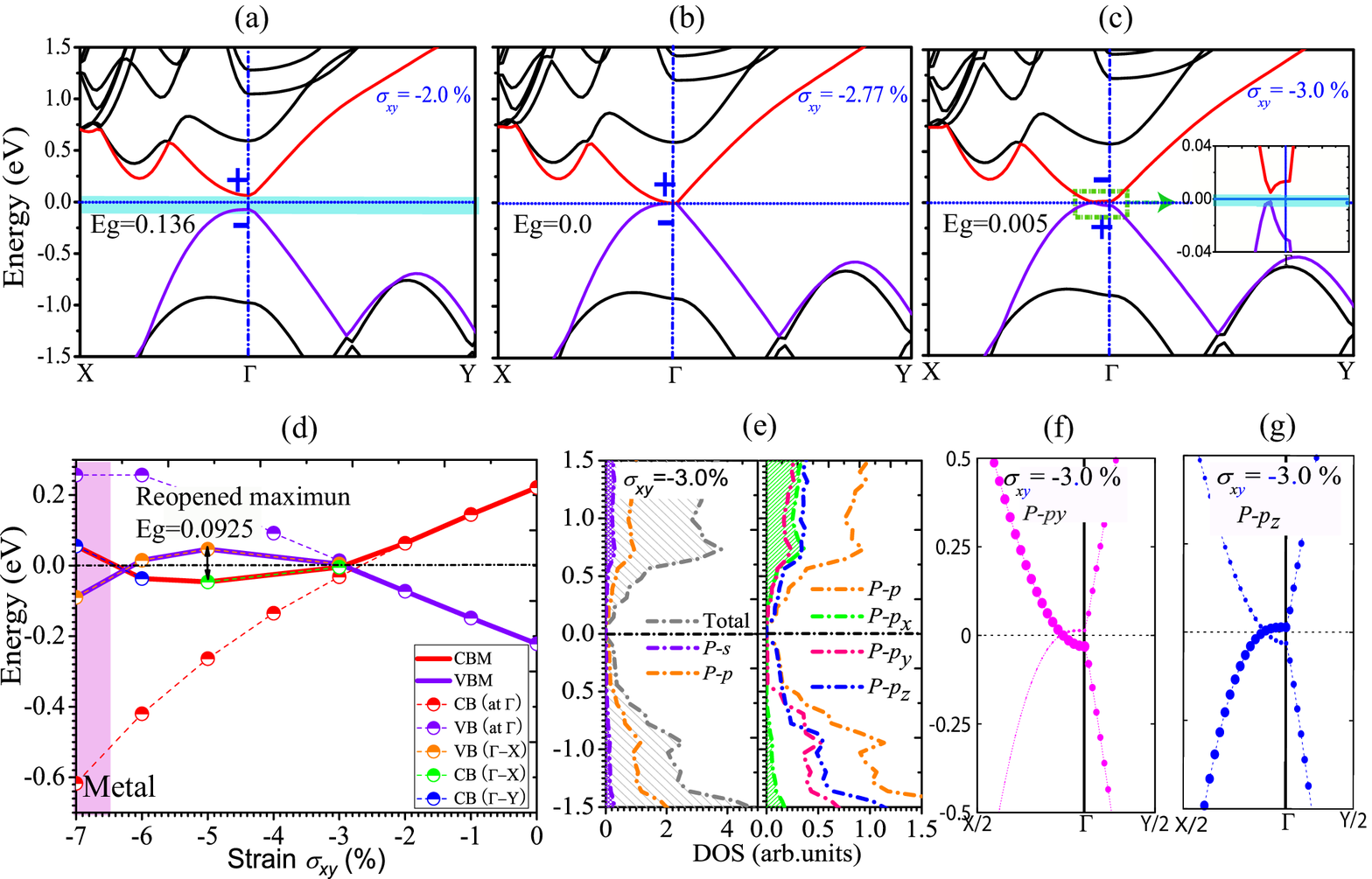,width=16cm}
    \end{center}
    \label{fig:CN}
\end{figure}

\newpage
\begin{figure}
    \begin{center}
        \epsfig{file=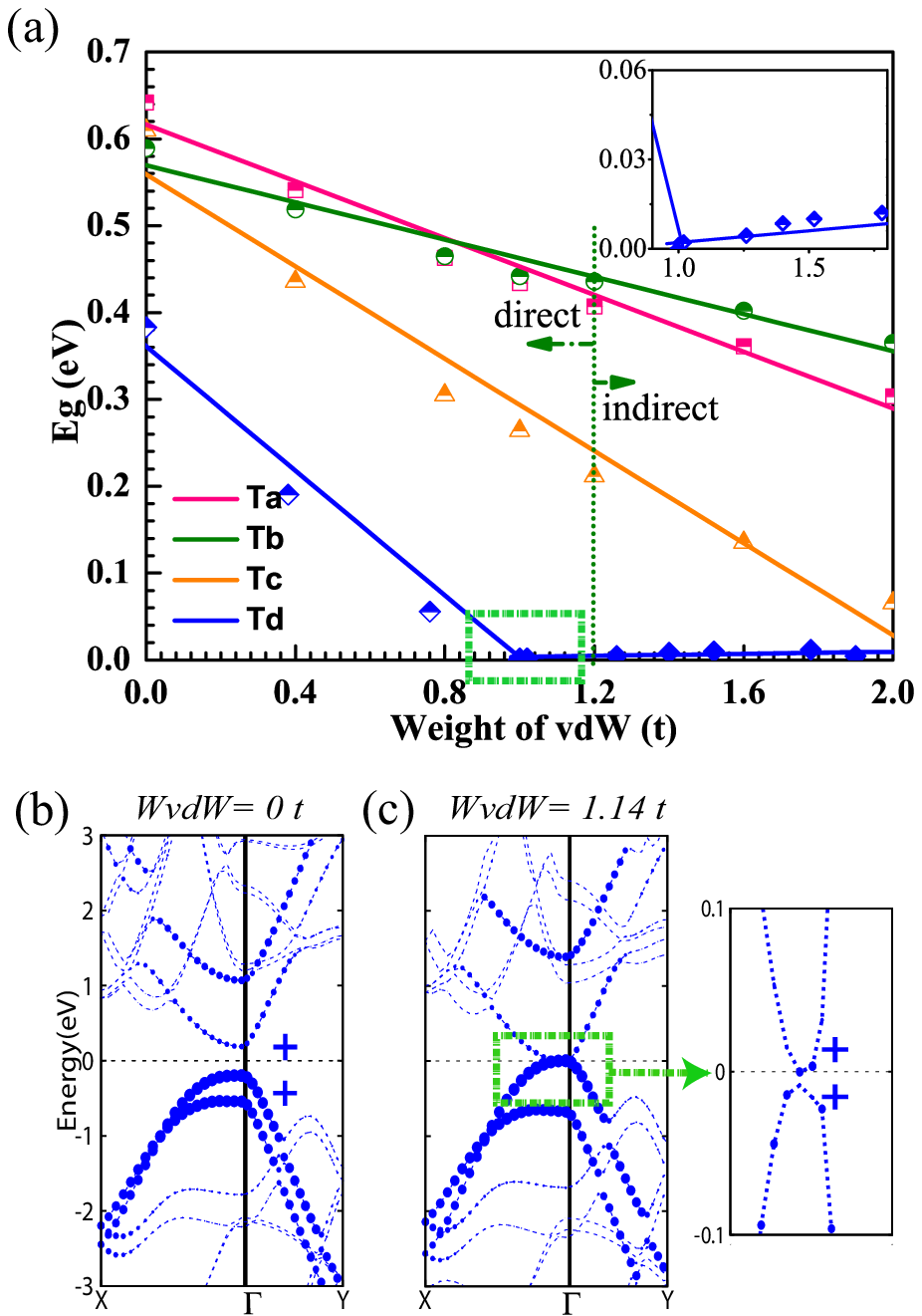,width=10cm}
    \end{center}
    \label{fig:CN}
\end{figure}

\newpage
\begin{figure}
    \begin{center}
        \epsfig{file=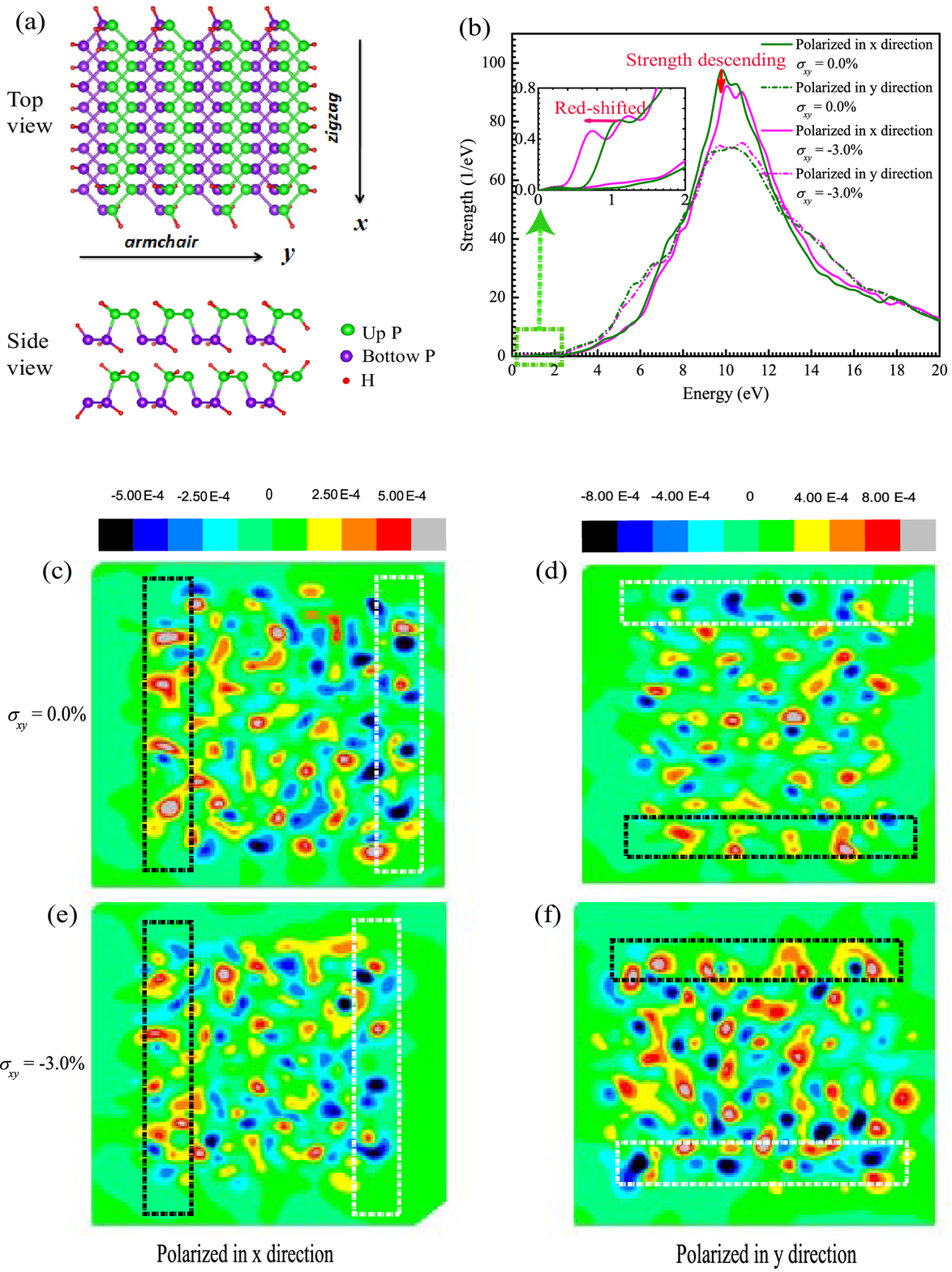,width=16cm}
    \end{center}
    \label{fig:CN}
\end{figure}
\newpage

\begin{table*}[!hc]
\centering \caption{ The calculated structural parameters of the
four different stacking orders, other previous theoretical
data\cite{JDXCZ} and the corresponding bulk experimental values
\cite{EXPT}, where $a$ ({\AA}) along x direction, $b$ ({\AA}) along
y direction are the lattice constants, $d_{int}$ ({\AA}) is the
smallest layer interval in the vertical direction, $R_{1}$ ({\AA}),
$R^{'}_{1}$ ({\AA}) are the in-plane bonds length, $R_{2}$ ({\AA})
is the out-plane bond length, $\alpha$ ($^{0}$) is the in-plane bond
angle and $\beta$ ($^{0}$) is the out-plane bond angle. $E_{coh}$
(eV/atom) is the cohesive energy with respect to isolated atoms, and
$\Delta$ $E_{coh}$(meV/atom)=$E_{coh}$-$E_{coh}$(Tb-stacked order)
is the relative stability of the different stacked bilayers with
respect to the Tb-stacked bilayer phosphorene.}
\begin{tabular}{cccccccccccccccc}
  \hline
  \hline
  &$a$&$b$&$d_{int}$&$R_{1}$&$R^{'}_{1}$&$R_{2}$&$\alpha$&$\beta$&$E_{coh}$&$\Delta$ $E_{coh}$\\
  \hline
  Ta         &$3.314$&$4.519$&$3.503$&$2.221$&$2.226$&$2.256$&$96.45$&$103.26$&$-3.6550$&$8.0$  \\
  Theory     &$3.326$&$4.550$&$3.495$&$2.243$&$2.235$&$2.283$&           &      &       &     \\
  Tb        &$3.319$&$4.505$&$3.108$&$2.223$&$2.226$&$2.253$&$96.55$&$103.17$&$-3.6630$&$0.0$ \\
  Theory     &$3.331$&$4.526$&$3.214$&$2.242$&$2.238$&$2.277$&           &    &         &       \\
  Tc         &$3.312$&$4.546$&$3.739$&$2.223$&$2.224$&$2.251$&$96.34$&$103.54$&$-3.6556$&$7.4$ \\
  Theory     &$3.324$&$4.535$&$3.729$&$2.238$&$2.236$&$2.274$&           &    &         &       \\
  Td         &$3.315$&$4.524$&$3.291$&$2.221$&$2.225$&$2.255$&$96.50$&$103.34$&$-3.6559$&$7.1$ \\
  Bulk(Exp)  &$3.314$&$4.376$&$3.503$&$2.224$&$2.244$&&$96.34$&$102.09$&         &       \\
  \hline
  \hline
\end{tabular}
\end{table*}

\newpage
\begin{table*}[!hc]
\centering
\caption{ Products of parity eigenvalues at four time-reversal invariant momentum for $0.0 \%$ and $-3.0 \%$ strain. Positive parity is denoted by $+$ while negative denoted by $-$. The resulting $Z_{2}$ values are shown.} 
\begin{tabular}{cccccccccccccccc}
  \hline
  \hline
  Strain  &  &  &  $\Gamma$ &  &  &$X$ &  &  & $Y$ &  &  &$ M$ &  &  & $\mathbb{Z}_{2}$\\
  \hline
  $0.0\%$  &  &  &  $+$  &  &  & $+$ &  &  & $+$ &  &  & $+$ &  &  & $0$\\
  $-3.0\%$ &  &  &  $-$  &  &  & $+$ &  &  & $+$ &  &  & $+$ &  &  & $1$\\
  \hline
  \hline
\end{tabular}
\end{table*}

\end{document}